\documentclass[12pt,preprint]{aastex}
\usepackage{emulateapj5}
\usepackage{apjfonts}
\usepackage{epsfig}
\usepackage{natbib}

\newcommand{\aox}{\ensuremath{\alpha_{\mathrm{ox}}}}

\newcommand{\chisq}{\ensuremath{\chi^2}}

\newcommand{\chandra}{\emph{Chandra}}

\newcommand{\etal}{et al.}

\newcommand{\flamb}{ergs s$^{-1}$ cm$^{-2}$ \AA$^{-1}$}

\newcommand{\feii}{\ion{Fe}{2}}

\def\gtrsim{\mathrel{\hbox{\rlap{\hbox{\lower4pt\hbox{$\sim$}}}\hbox{\raise2pt\hbox{$>$}}}}}

\newcommand{\fwhb}{\ensuremath{\mathrm{FWHM}_\mathrm{H{\beta}}}}
\newcommand{\gams}{\ensuremath{\Gamma_{\mathrm{s}}}}

\newcommand{\gamhr}{\ensuremath{\Gamma_{\mathrm{HR}}}}
\newcommand{\halpha}{H\ensuremath{\alpha}}

\newcommand{\hbeta}{H\ensuremath{\beta}}

\newcommand{\kms}{km~s\ensuremath{^{-1}}}

\newcommand{\lf}{\ensuremath{L_{5100}}}

\newcommand{\lledd}{\ensuremath{L_{\mathrm{bol}}/L{\mathrm{_{Edd}}}}}

\newcommand{\luv}{\ensuremath{L_{\rm{2500 \AA}}}}
\newcommand{\lx}{\ensuremath{L_{\mathrm{X}}}}

\newcommand{\mbh}{\ensuremath{M_\mathrm{BH}}}

\newcommand{\msigma}{\ensuremath{M_{\mathrm{BH}}-\sigmastar}}
\newcommand{\msun}{\ensuremath{M_{\odot}}}

\newcommand{\nh}{\ensuremath{N_{\mathrm{H}}}}

\newcommand{\oiii}{[\ion{O}{3}]}

\newcommand{\rosat}{\emph{ROSAT}}

\newcommand{\sigmastar}{\ensuremath{\sigma_{\ast}}}

\newcommand{\xmm}{{\it XMM-Newton}}

\def\lax{{$\mathrel{\hbox{\rlap{\hbox{\lower4pt\hbox{$\sim$}}}\hbox{$<$}}}$}}
\def\gax{{$\mathrel{\hbox{\rlap{\hbox{\lower4pt\hbox{$\sim$}}}\hbox{$>$}}}$}}

\slugcomment{Submitted to {\it The Astrophysical Journal} April 7, 2006.}
\shorttitle{X-ray Properties of Intermediate-mass Black Holes}
\shortauthors{GREENE \& HO}

\begin{document}

\title{X-ray Properties of Intermediate-mass Black Holes in Active Galaxies}

\author{Jenny E. Greene}
\affil{Harvard-Smithsonian Center for Astrophysics, 60 Garden St.,
Cambridge, MA 02138}

\author{Luis C. Ho}
\affil{The Observatories of the Carnegie Institution of Washington,
813 Santa Barbara St., Pasadena, CA 91101}

\begin{abstract}

We present a pilot study of the X-ray properties of intermediate-mass
($\sim 10^5-10^6$ \msun) black holes in active galaxies using the
\chandra\ X-ray telescope.  Eight of the 10 active galaxies are
detected with a significance of at least $3\,\sigma$, with X-ray
luminosities in the range $L_{0.5-2 \mathrm{keV}} \approx
10^{41}-10^{43}$ ergs s$^{-1}$.  The optical-to-X-ray
flux ratios are consistent with expectations, given the known
correlations between $\alpha_{\rm ox}$ and ultraviolet luminosity,
while a couple of objects appear to be anomalously X-ray weak.  The range of
0.5--2 keV photon indices we measure, $1 < \Gamma_{\mathrm{s}} < 2.7$,
is entirely consistent with values found in samples of more luminous
sources with more massive black holes.  Black hole mass evidently is
not a primary driver of soft X-ray spectral index.  On the other hand,
we do find evidence for a correlation between X-ray power-law slope and 
both X-ray luminosity and Eddington ratio, which may suggest that
X-ray emission mechanisms weaken at high Eddington ratio.  Such a
weakening may explain the anomalous X-ray weakness of one of our most
optically luminous objects.
\end{abstract}


\section{Introduction}

Astrophysical black holes (BHs) are typically found in two mass
ranges: stellar-mass BHs have masses $\sim 10$ \msun, and are the
remnants of massive star death, while supermassive BHs have masses in
the range $\sim 10^6-10^9$ \msun, and are a ubiquitous component of galaxy
bulges.  There are thus $\sim 5$ orders of magnitude in BH mass that
remains unexplored; BHs in this mass range have been dubbed
``intermediate-mass'' BHs, and their existence remains the subject of debate.  
In particular, there are anomalously luminous (``ultraluminous''; 
$L_{\mathrm{X}} \geq 10^{39}$ ergs s$^{-1}$) off-nuclear extragalactic
X-ray sources that may be BHs with masses of $100-1000$ \msun\ (see,
e.g., van der Marel 2004 for a review).  We have a complementary
program to find intermediate-mass BHs with masses of
$10^4-10^6$ \msun\ in active galactic nuclei (AGNs).

These objects are of interest not only because they begin to fill the
mass gap between supermassive and stellar-mass BHs, but also as
potential analogues of the primordial seeds of supermassive BHs.
Furthermore, the mergers of BHs with masses $\sim 10^5$~\msun\ are
expected to provide a strong signal for the {\it Laser Interferometric
Space Antenna} (e.g,~Hughes 2002).  The tight correlations between
supermassive BH mass (\mbh) and both the luminosity (Kormendy \&
Richstone 1995) and stellar velocity dispersion (the \msigma\
relation; Gebhardt \etal\ 2000; Ferrarese \& Merritt 2000; Tremaine
\etal\ 2002; Barth et al. 2005; Greene \& Ho 2006) of spheroids
suggest that BHs play an important role in the evolution of bulges.
On the other hand, very little is known about the prevalence of
nuclear BHs in late-type, bulgeless galaxies.  Unfortunately,
intermediate-mass BHs are difficult to find, because we are currently
unable to resolve the gravitational sphere of influence of a $\sim
10^5$~\msun\ BH outside the Local Group, and thus we cannot detect
them directly through resolved kinematics.  We are forced to rely on
indirect evidence of the presence of a BH from radiative signatures.
Two prototypical low-mass AGNs are known, in the late-type spiral NGC
4395 (Filippenko \& Ho 2003) and in the dwarf elliptical galaxy POX 52
(Barth \etal\ 2004).  Greene \& Ho (2004) systematically defined a
sample of similar objects using the First Data Release of the Sloan
Digital Sky Survey (SDSS; Abazajian \etal\ 2003).  Their sample of 19
galaxies forms the parent sample for the present paper, and throughout
we will refer to the sample using the identification numbers from
Greene \& Ho (2004).

The Greene \& Ho objects represent a very homogeneously selected sample
of BHs with low masses, and thus presents an ideal sample to investigate how 
the broad-band spectral properties of AGNs depend on BH mass.  The spectral 
properties of intermediate-mass BHs are important not
only for the insight they may provide into accretion processes, but
also to place observational constraints on the radiative properties of
``mini-quasars'' that may have contributed significantly to the
reionization of the Universe (e.g,~Madau \etal\ 2004).  We are thus in
the process of measuring the multiwavelength properties of the sample.
Using the Very Large Array, we found that the objects are very faint at 
in the radio (Greene et al. 2006).  Here we present the results of a pilot 
study to constrain the X-ray properties of the sample.

Throughout we assume the following cosmological parameters to calculate
distances: $H_0 = 100~h = 71$~\kms~Mpc$^{-1}$, $\Omega_{\rm m} = 0.27$,
and $\Omega_{\Lambda} = 0.75$ (Spergel \etal\ 2003).

\begin{figure*}
\vbox{ 
\vskip +0.2truein
\hbox{
\psfig{file=table1_v7.epsi,width=0.37\textwidth,keepaspectratio=true,angle=90}
}}
\end{figure*}

\vskip 5mm

\section{Observations and Data Analysis}

We observed 10 of the nearest intermediate-mass BHs from the sample
of Greene \& Ho (2004; see Table 1) with the Advanced CCD Imaging
Spectrometer (ACIS) onboard \chandra\ (Weisskopf \etal\ 1996).  The 5
ks observations were obtained during the Guest Observer Cycle 6,
between December 2004 and February 2006.  Images were obtained at the
aim point of the S3 CCD in faint mode.  In order to mitigate the
effects of pile-up, we read out only 1/8 of the chip to yield a
minimum read-out time of 0.4~s.  The effective exposure times ranged
from 5.01 to 5.67 ks.

All analysis was performed using the standard Level 2 event file
processed by the \chandra\ X-ray Center, which has cosmic ray
rejection and good time interval filtering included.  We use {\it
wavdetect} within CIAO ({\it Chandra} Interactive Analysis of
Observations) to extract positions for each source, in order to verify
that they are indeed the program objects.  Of the 10 objects
observed, eight are detected with a significance of $3\, \sigma$ or
greater (Table 1), where $\sigma$ is measured from the background
within the extraction aperture.  For the detected objects the SDSS and
\chandra\ positions agree within 0\farcs1 for all but GH10 and GH11,
which agree within 1\arcsec\ (these are the faintest detections).  The
on-axis point-spread function of \chandra\ contains 95\% of the
encircled energy within 1\arcsec.  We therefore extract counts from a
2\arcsec\ radius around each source, in the soft (0.5-2 keV; $C_{\rm
s}$) and hard (2--8 keV; $C_{\rm h}$) bands, while background rates
are calculated using annuli of inner radius 7\arcsec\ and outer radius
15\arcsec.  The task {\it dmextract} within CIAO is used to extract
the background-corrected count rates.  The background rates within
this aperture are truly negligible, consisting of fewer than 1 count
in each energy range over the total integration time.  We verify that
there is no extended emission by repeating the extraction for
1\arcsec\ and 3\arcsec\ extraction radii, and in all cases the
extractions are consistent with $95\%$ of the energy falling within
the 1\arcsec\ radius.  In GH02 we see marginal evidence for extended
emission; there is a $10\%$ increase in flux between the 1\arcsec\ and
2\arcsec\ extractions, but the result is only marginally significant.

\subsection{Hardness Ratios}
Only five objects are bright enough ($> 200$ cts) to perform proper
spectral fitting.  The ``hardness ratio'' [$H\,\equiv$ ($C_{\rm
h}-C_{\rm s}$)/($C_{\rm h}+C_{\rm s}$); Table 1] provides a crude
estimate of the spectral shapes of all the detected targets.  However,
we would like to derive a photon index over 0.5--2 keV (\gams; $N(E)
\propto E^{-\Gamma_{\mathrm{s}}}$) for each target in a uniform
manner.  We therefore follow Gallagher \etal\ (2005) and use the
observed hardness ratios combined with the instrumental response to
infer \gamhr\ for each source.  The instrumental response is expressed
in two matrices, the auxiliary response file (ARF) and the
redistribution matrix file (RMF), which we compute for each
observation using the task {\it psextract} in CIAO.  The ARF describes
the energy-dependent modifications to an input spectrum due to the
effective area and quantum efficiency of the telescope, while the RMF
modifies the input energy spectrum into the observed distribution of
pulse heights, due to the finite energy resolution of the detectors.
We use the spectral-fitting package XSPEC (Arnaud 1996) to generate
artificial spectra with known spectral slopes and Galactic absorption
(Table 1) over a range of soft photon indices $1 <$\gams$< 4$.  We
then ``observe'' the artificial spectra using the ARF and RMF
calculated for each observation, and measure a hardness ratio for each
input slope.  With this mapping between observed hardness ratio and
underlying slope, we infer \gamhr\ from hardness ratios, as shown in
Table 1.  Below we verify that \gamhr\ is consistent with more
detailed spectral analysis for the brightest objects (see Table 2),
which gives us some confidence that our approach is valid.  In the
interest of uniformity, however, when we compare distributions of
\gams, we will use only the values derived in this fashion.  We also
derive the fluxes shown in Table 1 using the measured count rates and
assuming \gamhr.

\begin{figure*}
\vbox{
\vskip -0.2truein
\hskip 0.in
\psfig{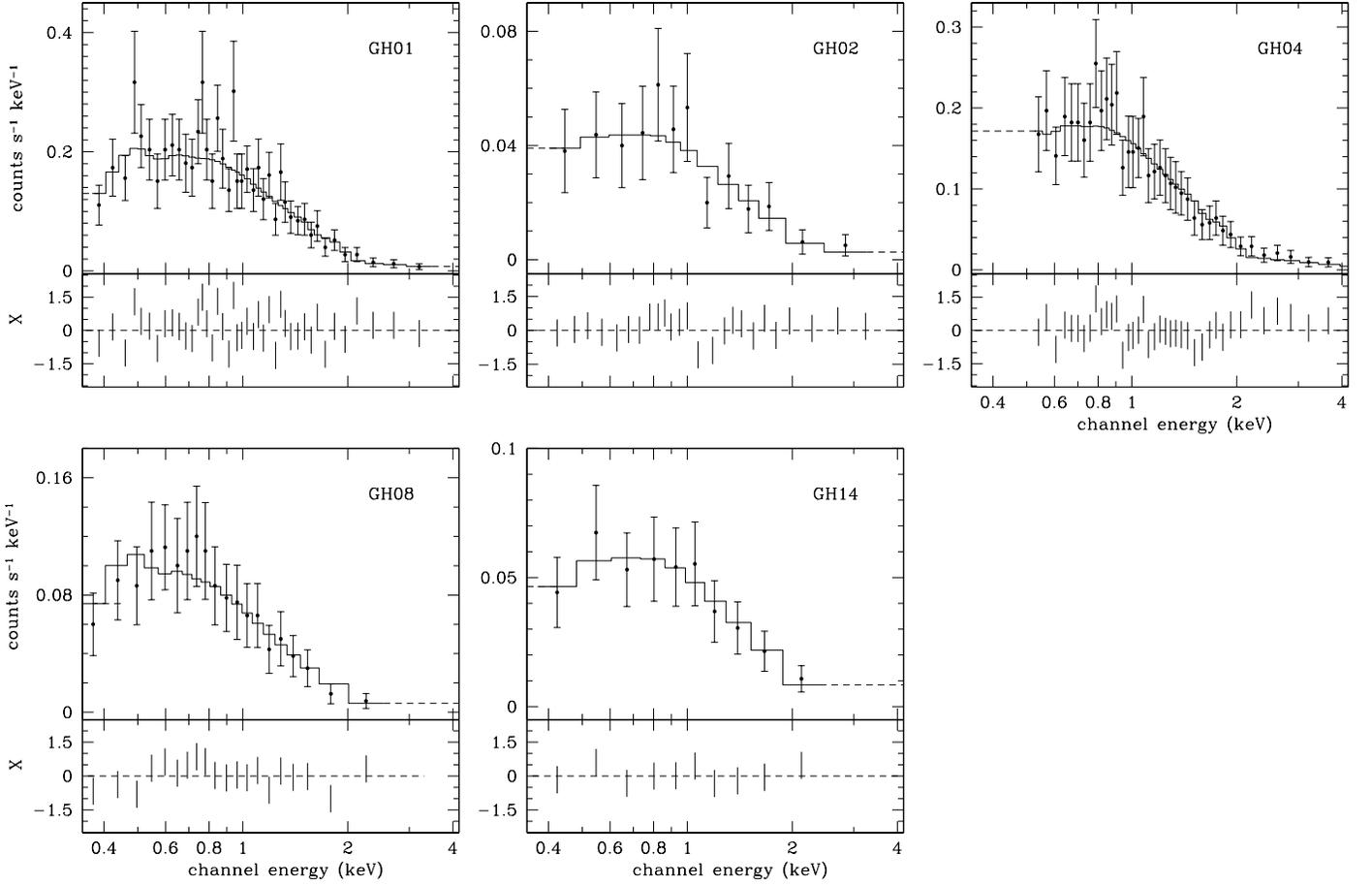}}
\vskip -0mm 
\figcaption[]{ 
Extracted spectra of GH01, GH02, GH04,
GH08, and GH14 from 0.3 to 4 keV, grouped to ensure a minimum of 20
counts in each energy bin.  Each spectrum is fit with an absorbed
power-law, where the absorption is fixed to the Galactic value (see
Table 1).  The best fits correspond to photon indices of \gams\ = $2.5
\pm 0.1$, $2.4 \pm 0.3$, $2.3 \pm 0.1$, $2.7 \pm 0.2$, and $2.2 \pm
0.2$, respectively.  The bottom panel in each plot shows the residuals
normalized by sigma ($\chi$).
\label{spec}}
\end{figure*}
\vskip 0mm

\begin{figure*}
\vbox{
\vskip -0.1truein
\hskip 0.5in
\psfig{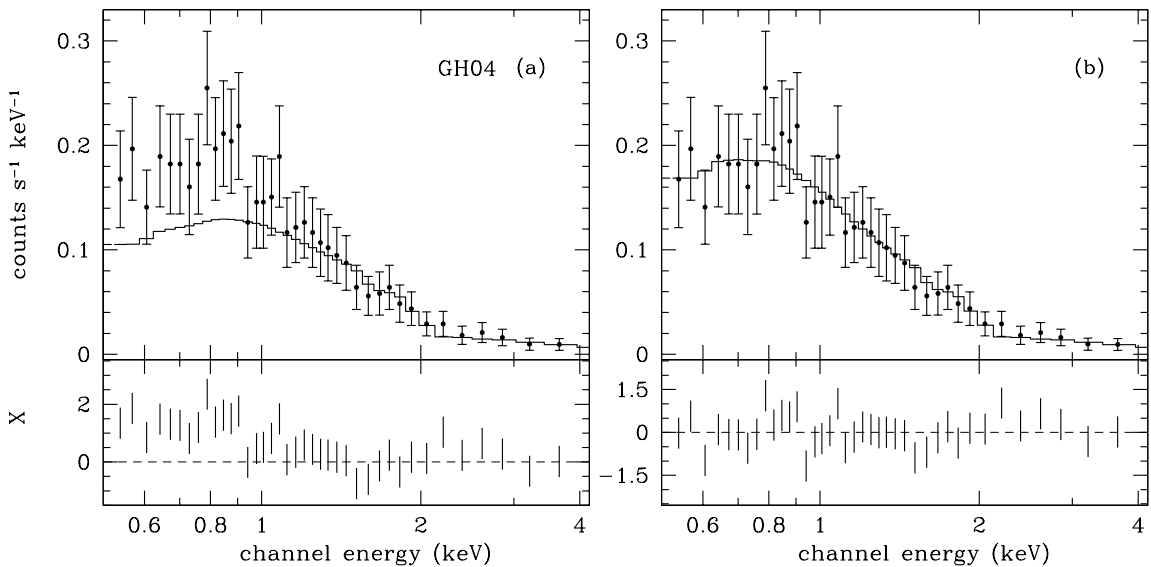}}
\vskip -0mm
\figcaption[]{
({\it a}) Spectral fit to GH04 using a single absorbed power-law,
restricted to energies $1.5-4$ keV.  Extraction and binning as in
Fig. 1, with absorption fixed to the Galactic value.  The best-fit
photon index is $\Gamma_{\rm s} = 1.9 \pm 0.5$; however, there is
clearly an excess at low energies.  ({\it b}) Same as in ({\it a}),
but with the power-law component fixed to a photon index of
$\Gamma_{\rm s} = 1.9$ and an additional blackbody component (in the
rest frame of the AGN) included.  A blackbody component with a
best-fit temperature of $kT = 0.15 \pm 0.24$ keV accounts for $40\%$
of the flux at 0.6 keV.
\label{gh04}}
\end{figure*}
\vskip 5mm

\hskip 0.0in
\psfig{file=table2_v5.epsi,width=0.4\textwidth,keepaspectratio=true,angle=0}
\vskip 2mm

\vskip 5mm

\subsection{Spectral Fitting}

Five (GH01, GH02, GH04, GH08, and GH14) of the 10 observed targets
had a sufficient total number of counts to enable a reliable spectral
fit.  First, the observations were binned to contain no fewer than 20
counts per bin, such that the \chisq\ statistic is valid.  Quoted
errors are for $90\%$ confidence in a single parameter.  The spectra
were then extracted using the aperture (2\arcsec) defined for the
aperture photometry. Since the background within this aperture is
neglible, no background subtraction is done.  We limit our attention
to the 0.3--5 keV range to avoid detector-related uncertainties at the
lowest energies and because there is virtually no signal above 5 keV.

We fit the spectra within XSPEC.  All of the spectra are well fit by a
simple absorbed power-law model, with the absorption fixed to the
Galactic value (Dickey \& Lockman 1990).  We show these fits in Figure
1 and Table 2.  This very simple model results in reasonable values of
\chisq\ in all cases, which suggests that we cannot justify additional
components.  Note also that we find good agreement between \gamhr\ and
\gams\ in all cases.  The fits with only Galactic absorption are
reasonable, but for completeness we have also performed fits with the
absorption allowed to vary.  In all cases, the derived values of
\gams\ and \nh\ agree within the uncertainties, while the goodness-of-fit 
is not significantly improved.  We therefore find no compelling
evidence for intrinsic neutral absorption.  Since the effective area
of the S3 chip peaks between 1 and 2 keV, with such short exposures we
cannot constrain well the hard ($\geq 2$ keV) continuum shapes.  If the
spectra are actually composed of multiple components, such as a
thermal excess at soft energies and a power-law at higher energies, we
have neither the required spectral coverage nor the needed depth to
adequately model these components.

However, significant soft excesses are a distinct possibility; the
soft component can dominate the spectrum up to energies as high as 1
keV (Brandt \etal\ 1997; Leighly 1999b).  For this reason, we
investigate whether we can place any constraints on the presence of a
soft excess.  It is common to model the soft excess as a blackbody
component (e.g., Leighly 1999b), and so we attempt to fit each
spectrum with a single blackbody component to investigate whether our
spectra are dominated by a soft excess component.  In all cases, the
blackbody provides a poor fit to the data, with significant excesses
at both low and high energies, and with unnaturally high energies of
$kT \approx 0.26$ keV, significantly higher than the typical
temperatures of $kT \approx 0.15$ keV seen in the soft X-ray excess of
both narrow-line Seyfert 1 (NLS1) galaxies (Leighly 1999b) and
Palomar-Green (PG; Schmidt \& Green 1983) quasars (Gierli\'{n}ski \&
Done 2004).  Unfortunately, we do not have sufficient counts to
warrant a multi-component fit for objects other than GH01 and GH04.
For these objects, we adopt a two-step procedure.  We first fit the
power-law component using only data $\geq 1.5$ keV, using the
extraction regions and binning as above, and fixing the absorption to
the Galactic value.  We then remodel the entire spectrum with the
derived power-law and an additional blackbody component at the
redshift of the AGN.  In the case of GH01, the putative extra
blackbody component would account for only $10\%$ of the flux at 0.6
keV.  The situation is different for GH04.  In this case, the best-fit
spectral index is marginally flatter, with \gams$= 1.9 \pm 0.5$.  As
is apparent from Figure 2{\it a}, when the model is extrapolated to
lower energies there is a significant excess above the simple power
law.  We then model the full spectrum (0.3--5 keV) with the power law
fixed and an additional blackbody component added in the rest-frame of
the AGN.  The combined fit (Fig. 2{\it b}) is quite reasonable,
although it is statistically indistinguishable from the single
power-law fit.  The blackbody component has a temperature of $kT =
0.15 \pm 0.24$ keV and accounts for $\sim 40\%$ of the flux at 0.6
keV.  Apparently the data are completely consistent with a soft excess
in GH04.  Because of the decreased sensitivity of \chandra\ above 2
keV, however, we cannot place strong constraints on the spectral shape
at higher energies, and thus can only roughly decompose the hard and
soft shapes.

\section{Results}

\subsection{Comparison with Narrow-line Seyfert 1 galaxies}

The Greene \& Ho (2004) sample occupies a unique regime in terms of BH
mass and Eddington ratio, and thus the distribution of X-ray
properties should provide interesting new constraints on physical
models of X-ray emission in AGNs.  The sample properties are, however,
well-bracketed at low mass by the prototypical intermediate-mass BHs
NGC 4395 and POX 52, and at high masses by the subclass of AGNs known
as NLS1 galaxies.  NLS1 galaxies were originally identified on the
basis of unusually narrow, but still kinematically distinctly broad
permitted lines---specifically \fwhb\ $< 2000$ \kms.  By this
definition, the Greene \& Ho sample, NGC 4395 and POX 52, all qualify
as NLS1s.  In general, however, NLS1s are also characterized by high
\feii/\hbeta\ and low \oiii/\hbeta\ ratios\footnote{Here \hbeta\
refers to the flux of the narrow component of the line.} in their
optical spectra (Osterbrock \& Pogge 1985).  They have since been
found to have very uniform X-ray properties, including a soft X-ray
excess (e.g.,~Boller \etal\ 1996), steep X-ray spectra (e.g.,~Leighly
1999b; Grupe \etal\ 2004), and extreme variability in the X-rays
(e.g,~Leighly 1999a).  Currently the best explanation for NLS1
properties is that they contain low-mass BHs\footnote{Collin \etal\
(2006) revisit the possibility that some NLS1s are most consistent
with having larger BH masses and a high inclination and suggest that
the Williams \etal\ sample is dominated by such objects.  While we
cannot rule out this possibility, the observation that the Greene \&
Ho sample obeys the low-mass extrapolation of the \msigma\ relation
(Barth \etal\ 2005; Greene \& Ho 2006) suggests that our objects are
truly low-mass BHs, as does the finding that the host galaxies
themselves are low-luminosity, late-type systems (Greene \& Ho 2004,
based on SDSS images; J. E. Greene, A. J. Barth, \& L. C. Ho, in
preparation, based on {\it Hubble Space Telescope} images).}
radiating at a high fraction of their Eddington rates (e.g.,~Pounds
\etal\ 1995).  In this picture, the broad lines are rather narrow due
to the low BH mass (and thus small virial velocities in the broad-line
region).  The characteristic strong \feii\ and weak \oiii\ emission
are thought to be generic spectral characteristics of AGNs in a high
accretion state (Boroson \& Green 1992; Boroson 2002).  NLS1s are also
characteristically radio-quiet, much like Galactic stellar-mass BHs in
a high accretion state (Greene \etal\ 2006; McClintock \& Remillard
2006).  The Greene \& Ho objects are technically NLS1s, based on the
linewidth criterion, and they are selected to have low masses.  They
also appear to be radiating close to their Eddington luminosities, and
they are uniformly radio-quiet (Greene \& Ho 2004; Greene \etal\
2006).  However, in terms of optical properties, the \feii\ and \oiii\
strengths of the Greene \& Ho sample span a larger range than typical
NLS1s (Greene \& Ho 2004).  We are now able to compare the X-ray
properties, particularly the spectral shapes and broader spectral
energy distributions, of this sample with NLS1s in general.

Prototypical NLS1 samples are characterized by an extreme soft X-ray
excess (e.g.,~Boller \etal\ 1996), which can dominate the spectrum
below $\leq 1$ keV (e.g.,~Brandt \etal\ 1997; Leighly 1999b).  While
GH04 may have a weak soft excess, we do not find compelling evidence
for a strong soft excess in our sample.  Only two objects (GH07 and
GH08) have notably steep soft spectral slopes compared to \gams\
$\approx$ 2.5 typical of AGNs (Yuan \etal\ 1998).  This finding is in
keeping with the results of Williams \etal\ (2004) that optically
selected NLS1s need not have steep soft X-ray photon indices.  With a
\fwhb\ of 1500 \kms\ (Filippenko \& Ho 2003) and a soft X-ray slope of
\gams\ $\approx 0.9$ (Lira \etal\ 1999; Moran \etal\ 1999), NGC 4395
perhaps provides the most dramatic example that low BH mass (or small
\fwhb) does not guarantee a steep soft slope.  However, the flat slope
of NGC 4395 may be misleading, since it probably results partly from
the presence of a warm absorber (e.g.,~Crenshaw \etal\ 2004).  While
the hard spectral slope of NGC 4395 is also quite flat ($0.6 < \Gamma_{\rm h}
< 1.7$), it is also observed to vary considerably, possibly due to
variations in the intrinsic absorption (Iwasawa \etal\ 2000; Shih
\etal\ 2003; Moran \etal\ 2005).  In contrast, we do not see evidence
for significant absorption in our spectra (at least from neutral
material).

Many NLS1 spectra are also found to have complicated features around 1
keV (e.g.,~Brandt \etal\ 1994; Leighly \etal\ 1997; Fiore \etal\ 1998;
Turner \etal\ 1998; Nicastro \etal\ 1999).  These features have been
explained in the literature as either absorption edges from highly
ionized oxygen, with blueshifts of $0.2c-0.6c$ (Leighly \etal\ 1997),
or resonant absorption lines (e.g.,~predominantly Fe L; Nicastro
\etal\ 1999).  More recently, Crummy \etal\ (2005, 2006) find that the
1 keV features result naturally in relativistically blurred
photoionized emission from the accretion disk (e.g.,~Ross \& Fabian
2005).  If the first explanation is correct, then these systems are
able to drive highly ionized outflows.  We do not have the energy
resolution nor the sensitivity to detect such features with high
confidence, but we note that all of our spectra show interesting
anomalies at $\sim 0.8-1$ keV.  We may be seeing the \ion{O}{7}
absorption edge at 0.74 keV, or we may be seeing a dip at $\sim 1$
keV, as seen by the authors above.  We note, too, that Williams \etal\
(2004) see a similar feature in their brightest object, SDSS
J1449+0022.

A final important characteristic of NLS1s in the X-rays is their
impressively rapid temporal variability (Boller \etal\ 1996).  The
variability amplitude appears to be correlated with soft excess
(Leighly 1999a, 1999b).  NGC 4395 is extremely variable; it was seen
to increase by a factor of 10 in $< 2000$~s.  Unfortunately, we cannot
constrain the variability properties of our sample from such short
observations.  However, we can look for long-term variability by
comparing our fluxes with those seen by the \rosat\ All-sky Survey
(RASS).  GH07, GH08, and GH14 were all detected by the RASS, and are
consistent with only small-amplitude ($< 50\%$) variability over a
$\sim$10 yr timescale.  In contrast, GH01 is $\sim 5$ times brighter
than in the RASS observation.  Also, GH04 was undetected by the RASS
but very clearly detected by \chandra, which is consistent with a
factor of $\sim 2$ variability.  Interestingly GH04 is our best
candidate for significant optical variability as well (T. Morton
\etal, in preparation).

\subsection{Broad-band Spectral Properties}

We turn briefly now to the broader spectral energy distributions of
this sample.  As noted above, the Greene \& Ho sample is extremely
radio-quiet, which is very similar to NLS1s in general (Greene \etal\
2006).  We now examine the ratio of X-ray to optical flux for this
subsample, using \aox, the slope of a hypothetical power law extending
from 2500 \AA\ to 2 keV
\hskip -0.1in
\psfig{file=alpha_ox_2arcsec_gh1.epsi,width=0.42\textwidth,
keepaspectratio=true,angle=0}
\vskip -0mm
\figcaption[]{
The spectral index \aox\ compared to the monochromatic 2500 \AA\ luminosity,
which is in units of \flamb.  {\it Filled circles}\ are the objects from this
study, for which $L_{2500 \AA}$ is derived from the power-law continuum fits of
Greene \& Ho (2005).  For comparison, we also plot the Williams \etal\ (2004)
sample ({\it open circles}), the PG NLS1s ({\it open triangles}), and
NGC 4395 ({\it asterisk}), as observed with \chandra\ (Moran \etal\ 2005).
The solid line represents the best fit $L_{2500 \AA}$-\aox\ relation of
Strateva \etal\ (2005), while the dashed lines bracket 1 $\sigma$ about the
best fit.
\label{aox}}
\vskip 5mm
\noindent
(e.g.,~Tananbaum \etal\ 1979).  We adopt the definition of Strateva
\etal\ (2005): \aox\ $\equiv -0.3838$log ($f_{2500
\mathrm{\AA}}/f_{\mathrm{2 keV}}$).  The flux density at 2500 \AA,
$f_{2500 \rm{\AA}}$, is calculated using the 5100 \AA\ continuum flux
densities and optical power-law continuum slope measurements from
Greene \& Ho (2005), in which the total continuum of each source was
modeled as a combination of host galaxy emission, AGN power-law
continuum, and \feii\ pseudo-continuum.  There is a well-known
correlation between \aox\ and continuum luminosity, typically measured
at 2500 \AA\ (e.g.,~Avni \& Tananbaum 1982; but see Bechtold \etal\
2003).  We have plotted \aox\ against \luv\ in Figure 3, with the
derived relation of Strateva \etal\ (2005).  GH14 and GH19 have
particularly strong galaxy continua and so their continuum slopes and
luminosities are not well constrained.  We also include the Williams
\etal\ (2004) sample of NLS1s with relatively weak X-ray luminosities
and the subset of PG quasars that qualify as NLS1s according to the
\fwhb\ criterion, with \fwhb\ taken from Boroson \& Green (1992).  Our
sample as a whole agrees reasonably well with a low-luminosity
extrapolation of the Strateva \etal\ relation; based on the median
luminosity log~\luv\ = 27.7, the relation predicts \aox\ of $-1.1 \pm
0.5$, which is consistent with the observed value for the majority of the
objects.  However, there are two possible outliers, GH05 and
GH11. (Recall that we are plotting the $3~\sigma$ upper limit for
GH05).  GH05 has optical properties very similar to NLS1s in general,
particularly a large \feii/\hbeta\ and small \oiii/\hbeta\ ratio.
These are properties typically associated with high Eddington ratios
(e.g.,~Boroson 2002).  On the other hand, GH11 has no apparent \feii\
lines, and it has a high \oiii/\hbeta\ ratio.

\begin{figure*}
\vbox{ 
\vskip -0.1truein
\hskip 0.in
\psfig{file=gamma_mass_ledd.epsi,width=0.35\textwidth,keepaspectratio=true,angle=-90}}
\vskip -0mm
\figcaption[]{
({\it a}) \gams(0.5--2 keV) vs. \mbh.  Observations from this paper
({\it filled circles}) fill in the low-mass region of this diagram,
and demonstrate that \gams\ is not determined by \mbh.  For comparison
we include the optically selected samples of Williams \etal\ (2004;
{\it open circles}) and the PG quasars observed with \xmm\ from
Porquet \etal\ (2004; {\it open squares}).  NGC 4395 is plotted as an 
{\it asterisk}.  ({\it b}) \gams\ vs. $L_{\rm 0.5-2 keV}$.  
({\it c}) \gams\ vs. \lledd.  We have
estimated \lledd\ from the optical luminosity alone, assuming $L_{\rm
  bol} = 9L_{5100 \AA}$, in order to avoid potential secondary
correlations between X-ray luminosity and slope, introduced, for
instance, by our increased sensitivity to soft sources.
\label{gamma}}
\end{figure*}
\vskip 5mm

\section{Physical Insights}

The primary goal of this study is to investigate the dependence of
X-ray properties on BH mass, since this AGN sample was uniformly
selected to have the lowest BH masses known.  In particular, we can
investigate the degree to which characteristic NLS1 properties are
driven by the BH mass.  Boller \etal\ (1996) found that \fwhb\ is
correlated with the soft X-ray slope and suggested that low BH mass
may be a necessary condition for steep soft photon indices (see also
Puchnarewicz \etal\ 1992; Laor \etal\ 1994; Wang \etal\ 1996).
Porquet \etal\ (2004) make a similar claim based on \xmm\ observations
of PG quasars.  However, our sample covers a range of
$600\leq$\fwhb$\leq 1800$ \kms\ ($10^5<$\mbh/\msun$<10^{6.5}$) and $1
\leq$\gams$\leq 3$.  We (and Williams \etal\ 2004) thus fill the
regions avoided by the Boller \etal\ sample.  Our mean \gamhr$=2.1 \pm
0.5$ (note that this mean neglects the non-detections) is, if
anything, {\it flatter}\ than \gams\ $=2.56 \pm 0.44$ measured in PG
quasars from 0.3--2 keV with \xmm\ by Porquet \etal\ (2004) or \gams\
$= 2.58 \pm 0.05$ measured for general \rosat\ samples (Yuan \etal\
1998).  We illustrate this in Figure 4{\it a}, where we plot \gams\
against \mbh\ for our sample, and the samples of Williams \etal\ and,
in order to increase the dynamic range in BH mass and luminosity, the
radio-quiet\footnote{Radio-loudness is defined as $f_{6
\mathrm{cm}}/f_{4500 \AA} > 10$.} sample of PG quasars with \xmm\
observations presented by Porquet \etal\ (2004).  The masses for the
Williams \etal\ sample were derived using the virial scaling relations
of Kaspi \etal\ (2000), while the BH masses for the Porquet \etal\
sample come from Ho (2002), using the same scaling relation.  The
masses in our sample are derived using a similar scaling relationship
based on the \halpha\ luminosity and linewidth (Greene \& Ho 2005), as
presented in Greene \& Ho (2006).  Finally, we also include NGC 4395,
using the \gams\ measured with \rosat\ by Moran \etal\ (1999) and the
reverberation-mapped mass from Peterson \etal\ (2005), as adapted by
Greene \& Ho (2006).  Clearly low BH mass is not a sufficient
condition for steep soft X-ray slopes.  The non-parametric Kendall's
$\tau =0.18$ with a probability $P_{\mathrm{null}}=0.08$ of no
correlation.  However, we note that while we have populated the region
of low BH mass (or narrow \hbeta) and shallow \gams, there is still an
unpopulated region of high BH mass (or broad \hbeta) and steep soft
\gams, as originally noted by Boller et al.  This may indicate that
mass still plays a secondary role in determining the spectral shape.
On the other hand, it may indicate a selection effect, in that
high-accretion rate, high-mass BHs are scarce in the local Universe,
while at higher redshift the relevant energy range is no longer
available.

Given that \mbh\ does not appear to be the main driver of soft X-ray
photon index, we seek correlation with other parameters.  In
particular, we consider both the soft (0.5--2 keV) X-ray luminosity,
$L_{\rm 0.5-2 keV}$, and the Eddington ratio, \lledd\ (Figs. 4{\it
b,c}).  The X-ray luminosities are derived from \chandra\ observations
for our sample and that of Williams et al., while they are measured from
0.3 to 2 keV with \xmm\ for the Porquet \etal\ sample, and converted to
a common bandpass of 0.5--2 keV using the spectral slopes derived by
Porquet et al.  The Eddington ratios are derived using \lf\ and
assuming $L_{\rm{bol}} = 9 L_{5100 \AA}$ (e.g., Kaspi et al. 2000).
We have used the optical luminosities here to avoid a potentially
spurious correlation between \gams\ and \lx\ which may arise because
we are more sensitive to soft sources with \chandra.  From this
(incomplete and heterogeneous) sample, we find the strongest
correlation between \aox\ and X-ray luminosity ($\tau = 0.46,\,
P_{\mathrm{null}}=6 \times 10^{-6}$), subject to the caveat mentioned
above, while the correlation with Eddington ratio is also significant
($\tau = 0.30, \, P_{\mathrm{null}}=0.003$).

We are by no means the first to claim a correlation between X-ray
spectral slope and X-ray luminosity.  It has been found that \gams\
correlates with soft X-ray luminosity in NLS1s (Forster \& Halpern
1996; Williams \etal\ 2004), and more general AGN populations (Lu \&
Yu 1999), while $\Gamma_{\mathrm{h}}$ (the photon index from 2--10
keV) has been found to correlate with hard X-ray luminosity as well
(Dai \etal\ 2004; Gierli{\'n}ski \& Done 2004; Porquet \etal\ 2004;
Wang \etal\ 2004).  Interestingly, individual objects have been
observed to obey a similar correlation as they vary (e.g.,~Chiang
\etal\ 2000; Petrucci \etal\ 2000; Vaughan \& Edelson 2001).
Furthermore, strong correlations between \gams\ and \feii\ strength
(or the \feii/\hbeta\ ratio) have long been known (e.g.,~Wilkes \etal\
1987; Shastri \etal\ 1993; Laor \etal\ 1994; Wang \etal\ 1996; but see
also Boroson 1989), which may support a trend between Eddington ratio
and \gams\ since \feii\ strength correlates with Eddington ratio
(e.g.,~Boroson 2002).  However, we are still wary that the presence or
absence of X-ray slope versus luminosity correlations may depend on
the particular sample chosen, since many people have {\it not} found
this trend (Leighly 1999b for NLS1s; George et al. 2000; Reeves \&
Turner 2000).  Clearly, a well-defined sample covering a complete
range in BH mass and luminosity using a single instrument is needed to
address this question definitively.  For the time being, we consider
the above trend suggestive.

Various groups have proposed explanations for the trend between hard
X-ray luminosity and hard photon index.  Dai \etal\ (2004) hypothesize that
their sample occupies a narrow range in Eddington ratio, with a wide
range in BH mass.  They invoke the disk-corona model of Haardt \&
Maraschi (1993), in which cool disk photons provide the seeds for
Compton cooling of the hard X-ray emitting corona, as well as
thermally reprocessing and reflecting some fraction of the hard X-ray
emission.  Haardt \etal\ (1997) show that within this model if the
optical depth of the corona is dominated by electron-positron pairs,
then there will be a correlation between 2--10 keV spectral shape and
luminosity.  On the other hand, Wang \etal\ (2004) find that the
fraction of bolometric luminosity emitted in the hard X-ray band decreases
with \lledd, and thus conclude that the hard X-ray emitting region
weakens at high Eddington ratio.  They propose that the coupling
between the disk and the corona depends on magnetic turbulence, which may
depend on the Eddington ratio, depending on the form of the magnetic
stress tensor.  Generalizing these models to the 
soft photon index is made complicated by the fact that the origin of
the soft X-ray excess is unknown.  Pounds \etal\ (1995)
proposed that low-mass BHs at high Eddington ratios would have a soft
thermal disk component observable in the \rosat\ band.  However, we
have seen that BH mass does not drive \gams, and furthermore,
Gierli{\'n}ski \& Done (2004) argue that the soft X-ray excess is too
hot by $\sim 0.1$ keV and (particularly) too constant in temperature
($\sim 0.15$ keV; see also Laor \etal\ 1994; Leighly 1999b) to arise
from the accretion disk.  They propose that relativistically blurred
absorption from a wind causes an apparent excess at soft energies.  In
this context, it is possible that a higher luminosity (and thus more
vigorous wind; e.g.,~Proga \etal\ 2000) may lead to a steeper apparent
slope.  On the other hand, Crummy \etal\ (2006) model the soft excess
using the ionized-reflection models of Ross \& Fabian (2005).  In this
case, the soft X-ray emission arises from photoionization of a highly
ionized, relativistically rotating inner accretion disk.  In this
model, the apparent soft spectral slope does steepen for steeper input
hard X-ray slopes.

In light of the finding of Wang \etal\ (2004) that the corona weakens
at high Eddington ratio, one would expect \aox\ to steepen with
increasing ultraviolet (UV) luminosity, particularly for a relatively
narrow range in \mbh.  However, we would then expect our low-mass,
low-luminosity, and high-\lledd\ sample to deviate significantly from
the relation derived for more massive BHs, which is not the case
(Fig. 3).  In fact, we do not see a significant correlation between Eddington
ratio and \aox\ in our sample; it spans more than an order of
magnitude in \lledd\ and yet is nearly constant in \aox.  Rather, our
observations support a scenario in which the bolometric importance of
the X-ray-emitting region is determined by the strength of the UV
continuum.  Disk-corona models (e.g.,~Haardt \& Maraschi 1993), in
which the soft disk photons cool the corona and the corona heats the
disk photons, naturally explain such a relationship.  Nevertheless,
\aox\ must vary with \lledd\ as well.  At very low accretion rates,
there is believed (and observed) to be a transition from an optically
thick, geometrically thin disk (Shakura \& Sunyaev 1973) to a
radiatively inefficient, optically thin and geometrically thick flow
(e.g.,~Ho 1999, 2005; Quataert 2001; Narayan 2005 and references
therein). If there is no optically thick accretion disk at all, and
thus no peak in the thermal disk emission in the far-UV, then \aox\
should flatten considerably (L. C. Ho, in preparation).  NGC 4395, for
instance, has both a very low Eddington ratio ($\sim 0.04$; Moran
\etal\ 1999) and an abnormally flat spectral slope (see Figure 3).

There is also some evidence that \gams\ may steepen considerably at
very high Eddington ratios.  Specifically, there is a subset of NLS1s
that appear to be intrinsically X-ray-weak (Leighly 2001; Leighly
\etal\ 2001; Williams \etal\ 2004).  While X-ray absorption can also
cause anomalously steep \aox\ values (e.g.,~Brandt \etal\ 2000;
Gallagher \etal\ 2002), there is no evidence for neutral absorption in
the X-ray spectra (or the UV spectra in the case of the Leighly
objects).  Extreme variability may explain the X-ray weakness in some
of these objects, but it cannot reasonably account for all of them.
Leighly \etal\ (2001) speculate that the X-ray weakness reflects a
truly weak corona, either because the accretion disk extends so close
to the BH that there is physically no space for a corona, or, like in
the Wang \etal\ (2004) picture, because the disk-corona energy
transport mechanism is quenched at high luminosity.  We
note, as an aside, that low-ionization broad absorption-line quasars,
which are believed to be in a high accretion state (e.g.,~Meier 1996;
Boroson 2002), are also proving particularly difficult to detect in
the X-rays (e.g.,~Green \etal\ 2001; Gallagher \etal\ 2006).  While
these objects may simply be Compton-thick, it is also possible that
they are intrinsically X-ray-weak.  In our own sample, GH11 has a flat
\gamhr, and is quite consistent with absorption driving the low X-ray
flux.  On the other hand, it would be very useful to obtain deeper
X-ray observations of GH05 in order to determine whether it is
intrinsically X-ray-weak or heavily absorbed.  The interpretation that
X-ray weakness is related to high Eddington ratio is quite
speculative, but could provide a powerful diagnostic to isolate
objects in a specific accretion regime.

\section{Summary}

We present a pilot study of the X-ray properties of intermediate-mass
BHs in active galaxies.  We have detected eight out of the 10 
objects surveyed with 5 ks \chandra\ observations.  From an analysis
of their hardness ratios, we find that the objects have a range in
0.5--2 keV photon index of $1<$\gams$<3$ [$N(E) \propto
E^{-\Gamma_{\mathrm{s}}}$], in keeping with general AGN samples
previously studied.  In the five objects for which there are sufficient
counts to extract spectra, there is no evidence for neutral
absorption, although we do see evidence for spectral complexity at
energies $\leq 1$ keV, which may indicate ionized absorption.  The
X-ray-to-optical flux ratios for the majority of the sample obey the
same correlation with optical/UV luminosity seen in the literature,
but there are two X-ray-weak sources.  Given the apparent correlation
between Eddington ratio and spectral slope, we speculate that X-ray
weakness in at least one of these sources may be related to a high
Eddington ratio.  We find that the soft X-ray slope correlates strongly 
with soft X-ray luminosity, but not with BH mass.  

This study demonstrates the feasibility of detecting $\sim
10^5-10^6$~\msun\ BHs with relatively short \chandra\ observations.  With
them, we have begun to probe the accretion properties of AGNs in an
unexplored mass regime.  It is our goal to build broad spectral energy
distributions for the entire sample, in order to properly measure
their bolometric luminosities.  We will then be in a good position to
search for mass-dependent spectral properties, as well as calibrate
various common average bolometric corrections (such as \oiii\
luminosity) for objects in this mass regime.  Finally, while we are
unable to address the variability properties of the sample with short
observations, there are strong indications that the X-ray variability
amplitude (e.g.,~O'Neill \etal\ 2005) and detailed power spectrum
(McHardy \etal\ 2005) can serve as indicators of the BH mass.
Extended X-ray observations of this sample would provide a powerful
new test of these important BH mass measurement techniques.

\acknowledgements
We are grateful for helpful assistance on observing from the 
\chandra\ X-ray Center and on data analysis from
J.~D.~Gelfand, R.~C.~Hickox, S.~Bogdanov, and E.~M.~Kellogg.  We 
thank I.~V.~Strateva for useful discussions.  
This research has made use of
data obtained from the High Energy Astrophysics Science Archive
Research Center (HEASARC), provided by NASA's Goddard Space Flight
Center.  L.~C.~H. acknowledges support by the Carnegie Institution of
Washington and by NASA grant SAO 06700600.

\end{document}